\documentclass[pra,twocolumn,amsmath,18pt]{revtex4-1}

\usepackage{amsthm}
\usepackage{amsmath}
\usepackage{amssymb}
\usepackage{braket}
\usepackage{bbold}
\usepackage{graphicx}
\usepackage{subfigure}
\usepackage{hyperref}
\usepackage{appendix}
\usepackage{tikz}
\usepgflibrary{shapes.geometric}
\usetikzlibrary{arrows}

\theoremstyle{definition}

\hypersetup{colorlinks=true, citecolor=blue, urlcolor=blue, linkcolor=blue}

\begin{document}
\title{General monogamy relations of the $S^{t}$ and $T^{t}_q$-entropy entanglement measures based on dual entropy}
\author{Zhong-Xi Shen$^1$}
\email{18738951378@163.com}
\author{Kang-Kang Yang$^1$}
\email{2220501004@cnu.edu.cn}
\author{Zhi-Xiang Jin$^2$}
\email{zxjin@dgut.edu.cn}
\author{Zhi-Xi Wang$^1$}
\email{wangzhx@cnu.edu.cn}
\author{Shao-Ming Fei$^1$}
\email{feishm@cnu.edu.cn}
\affiliation{$^1$School of Mathematical Sciences, Capital Normal University,
Beijing 100048, China\\
$^2$School of Computer Science and Technology, Dongguan University of Technology,
Dongguan 523808, China}

\bigskip

\begin{abstract}
Monogamy of entanglement is the fundamental property of quantum systems. By using two new entanglement measures based on dual entropy, the $S^{t}$-entropy entanglement and $T^{t}_q$-entropy entanglement measures, we present the general monogamy relations in multi-qubit quantum systems. We show that these newly derived monogamy inequalities
are tighter than the existing ones. Based on these general monogamy relations, we construct the set of multipartite entanglement indicators for $N$-qubit states, which are shown to work well even for the cases that the usual concurrence-based indicators do not work. Detailed examples are presented to illustrate our results.\\

\end{abstract}

\maketitle

\section{introduction}
As a fundamental issue of quantum  mechanics, quantum entanglement is the most important resource in quantum information processing \cite{JMHA2017,WMYX2018,HHGB2018,DFGR2017}. The characterization and quantification of entanglement is of vital significance. A variety of entanglement measures have been proposed from different perspectives to describe the degree of inseparability of multipartite quantum states, for instance, the concurrence \cite{Hill1997}, entanglement of formation  \cite{Bennett19963824}, R\'{e}nyi-$\alpha$ entropy entanglement \cite{HHH1996,Gour2007,Kim2010R}, Tsallis-$q$ entropy entanglement \cite{LV1998,Kim2010T}, and Unified-$(q,s)$ entropy entanglement \cite{KimBarry2011}. Recently, a new entanglement measure, called $S^{t}$-entropy entanglement, has been presented by adding its complementary dual part to the well-known von Neumann entropy, which can be viewed as a quantum version of entropy \cite{Yang2023}. Another new entanglement measure, $T^{t}_q$-entropy entanglement, is proposed in Ref.~\cite{Yang2023} based on the total entropy of Tsallis-$q$ entropy and its complementary dual. Both measures are analytically computable for any $N$-qubit states.

The monogamy of entanglement is a key property characterizing the entanglement sharability in multipartite quantum systems. Coffman, Kundu and Wootters~(CKW)  first characterized the monogamy of an entanglement measure $\mathcal{E}$ for three-qubit states $\rho_{ABC}$~\cite{CKW2000},
\begin{equation}\label{CKW}
\mathcal{E}(\rho_{A|BC})\geq \mathcal{E}(\rho_{AB})+\mathcal{E}(\rho_{AC}),
\end{equation}
where $\rho_{AB}={\rm tr}_C(\rho_{ABC})$, $\rho_{AC}={\rm tr}_B(\rho_{ABC})$ are the reduced density matrices of $\rho_{ABC}$, $\mathcal{E}(\rho_{A|BC})$ stands for the entanglement under the bipartition $A$ and $BC$. This relation is called monogamy of entanglement \cite{CKW2000,Terhal2004}. Later, Osborne and Verstraete extended this monogamy inequality to the squared concurrence for $N$-qubit systems \cite{T.J.Osborne}. Extensive researches have been conducted on the distribution of entanglement in multipartite quantum systems by employing various measures such as the squared entanglement of formation (EOF) \cite{Oliveira2014,Bai3,Bai2014}, the squared R\'{e}nyi-$\alpha$ entropy \cite{R2015}, the squared Tsallis-$q$ entropy \cite{Luo2016} and the squared Unified-$(q,s)$ entropy \cite{Khan2019}. Generally, monogamy inequalities depend on both detailed measures of entanglement and detailed quantum states. It has been shown that the squashed entanglement is monogamous for arbitrary dimensional systems \cite{Christandl2004}. Interestingly, a set of tight $\alpha$-th powers monogamy relations have been investigated for multi-qubit systems \cite{Zhu2014,Luo2015,Luo2016,JF2017,JF2018}. The traditional monogamy inequality (\ref{CKW}) provides a lower bound for ``one-to-group'' entanglement, i.e., the quantum marginal entanglement \cite{Walter2013}.

The monogamy inequality corresponds to a residual quantity~\cite{Zhu2014}, for example, the concurrence corresponds to the 3-tangle. The residual measure derived from the entanglement of formation is demonstrated to serve as an indicator for multi-qubit entanglement, capable of detecting all genuine multipartite entangled states \cite{Bai3}. These monogamy relations also play an important role in quantum information theory \cite{See2010}, condensed-matter physics \cite{Ma2011} and even black-hole physics \cite{Ve2013}.

The rest of this paper is organized as follows. In Sec.\ref{Sec1} and Sec.\ref{Sec2}, we review some background knowledge on entanglement measures that will be used in the main text, and establish two classes of tighter monogamy inequalities for $S^{t}$-entropy entanglement and $T^{t}_q$-entropy entanglement measures, respectively. In Sec.\ref{Sec3}, we investigate multipartite entanglement indicators based on two new monogamy relations for $N$-qubit states, together with detailed examples. We summarize our main results in Sec.\ref{Sec4}.

\section{Monogamy of $S^{t}$-entropy entanglement}\label{Sec1}
The $S^{t}$-entropy entanglement of a pure bipartite state $|\Phi\rangle_{AB}$ in $d\times d$ dimensional Hilbert space ${\cal H}_A\otimes {\cal H}_{B}$ is given by
\begin{eqnarray}
E_t(|\Phi\rangle_{AB})=\frac{1}{r}S^{t}(\rho_A),
\label{s1}
\end{eqnarray}
where $r=d\log_2d-(d-1)\log_2(d-1)$ is a normalization factor, $\rho_A={\rm Tr}_B(|\Phi\rangle_{AB}\langle\Phi|)$ is the reduced density operator with respect to the subsystem $A$, and $S^{t}(\rho)$ is the total entropy of a quantum state $\rho$ defined by
\begin{eqnarray}
S^{t}(\rho)=-{\rm{Tr}}[\rho\log_2\rho+(\mathbb{1}-\rho)\log_2(\mathbb{1}-\rho)],
\label{s2}
\end{eqnarray}
with $\mathbb{1}$ the identity matrix.
For a bipartite mixed state $\rho_{AB}$ in ${\cal H}_A\otimes {\cal H}_{B}$, the $S^{t}$-entropy entanglement is given via the convex-roof extension,
\begin{eqnarray}
E_t(\rho_{AB})=\inf_{\{p_i,|\Phi_i\rangle\}}\sum_ip_iE_t(|\Phi_i\rangle_{AB}),
\label{s3}
\end{eqnarray}
where the infimum is taken over all the possible pure state decompositions of $\rho_{AB}=\sum_ip_i|\Phi_i\rangle\langle\Phi_i|$ with $p_i\geq0$, $\sum_ip_i=1$.

In Ref.~\cite{Yang2023} the authors provide an analytic formula of the $S^{t}$-entropy entanglement for two-qubit systems based on concurrence. The concurrence of a bipartite pure state $|\Phi\rangle_{AB}$ is defined by \cite{Rungta2001},
\begin{eqnarray}
C(|\Phi\rangle_{AB})=\sqrt{2(1-{\rm Tr}(\rho^2_A))}.
\label{s4}
\end{eqnarray}
For mixed states $\rho_{AB}$, the concurrence is given by the convex-roof extension,
\begin{eqnarray}
C(\rho_{AB})=\inf_{\{p_i,|\Phi_i\rangle\}}
\sum_ip_iC(|\Phi_i\rangle_{AB}),
\label{s5}
\end{eqnarray}
where the infimum takes over all the possible pure-state decompositions of $\rho_{AB}$. In particular, for a two-qubit mixed state $\rho$ the concurrence has the analytic formula \cite{CKW2000},
\begin{eqnarray}
C(\rho)=\max\{0,\eta_1-\eta_2-\eta_3-\eta_4\},
\label{s6}
\end{eqnarray}
with $\eta_i$ the eigenvalues of the matrix $\sqrt{\rho(\sigma_Y\otimes\sigma_Y)\rho^* (\sigma_Y\otimes\sigma_Y)}$ in decreasing order, where $\rho^*$ is the complex conjugate of $\rho$ and $\sigma_Y$ is the standard Pauli operator.

Consider any ${\cal C}^2\otimes {\cal C}^d$ pure state $|\phi\rangle_{AB}$ in $\mathcal{H}_A\otimes\mathcal{H}_B$ with Schmidt form,
\begin{eqnarray}
|\phi\rangle_{AB}=\sqrt{\lambda_0}|0\rangle|\phi_0\rangle
+\sqrt{\lambda_1}|1\rangle|\phi_1\rangle,
\label{Schmidt}
\end{eqnarray}
where the subsystem $A$ is a qubit system, while the subsystem $B$ is a $d$ dimensional space, $|\phi_0\rangle$ and $|\phi_1\rangle$ are orthogonal states in $\mathcal{H}_B$, $\lambda_0$ and $\lambda_1$ are the Schmidt coefficients. From Eq.~(\ref{s1}) we have
\begin{eqnarray}
E_t(|\phi\rangle_{AB})=-\lambda_0\log_2\lambda_0-\lambda_1\log_2\lambda_1.
\label{hx1}
\end{eqnarray}
Besides, the concurrence of $|\phi\rangle_{AB}$ is given by
\begin{eqnarray}
C(|\phi\rangle_{AB})=\sqrt{2(1-{\rm{Tr}}(\rho^2_A))}=2\sqrt{\lambda_0\lambda_1}.
\label{C}
\end{eqnarray}

For any two-qubit pure state $|\phi\rangle_{AB}$ one has \cite{Yang2023},
\begin{eqnarray}
E_t(|\phi\rangle_{AB})=h(C(|\phi\rangle_{AB})),
\label{s7}
\end{eqnarray}
where $h(x)$ is an analytic function defined by
\begin{eqnarray}
h(x)
&=&-\frac{1+\sqrt{1-x^2}}{2}\log_2\frac{1+\sqrt{1-x^2}}{2}
\nonumber\\
& &
-\frac{1-\sqrt{1-x^2}}{2}\log_2(\frac{1-\sqrt{1-x^2}}{2}).
\label{s8}
\end{eqnarray}
Thus, we get a functional relation (\ref{s7}) between the concurrence and the $S^{t}$-entropy entanglement for any qubit-qudit pure state in $\mathcal{H}_A\otimes\mathcal{H}_B$.
It has been shown that the relation (\ref{s7}) holds also for two-qubit mixed states $\rho_{AB}$ \cite{Yang2023},
\begin{eqnarray}
E_t(\rho_{AB})=h(C(\rho_{AB})).
\label{s9}
\end{eqnarray}

The EOF is defined by \cite{Bennett19963824,Wootters1998},
\begin{eqnarray}
E_f(|\Phi\rangle_{AB})=-{\rm{Tr}}(\rho_A\log_2\rho_A)
\label{s10}
\end{eqnarray}
for any pure state $|\Phi\rangle_{AB}$ in ${\cal H}_A\otimes {\cal H}_{B}$, and
\begin{eqnarray}
E_f(\rho_{AB})=\inf_{\{p_i,|\Phi_i\rangle\}}\sum_ip_iE_f(|\Phi_i\rangle)_{AB}
\end{eqnarray}
for any bipartite mixed state $\rho_{AB}$, where the infimum takes over all the possible pure-state decompositions of $\rho_{AB}$.
It is shown in Ref.~\cite{Wootters1998} that $E_{f}(|\phi\rangle_{AB})=f(C^2(|\phi\rangle_{AB}))$ for any $2\otimes m $ ($m\geq2$) pure state $|\phi\rangle_{AB}$, and $E_{f}(\rho_{AB})=f(C^2(\rho_{AB}))$ for any two-qubit mixed state $\rho_{AB}$, where $f(x)$ is an analytic function defined by
\begin{eqnarray}
f(x)&=&-\frac{1+\sqrt{1-x}}{2}\log_2\frac{1+\sqrt{1-x}}{2}
\nonumber\\
& &
-\frac{1-\sqrt{1-x}}{2}\log_2(\frac{1-\sqrt{1-x}}{2}).
\label{E}
\end{eqnarray}
Thus the $S^{t}$-entropy entanglement reduces to EOF for two-qubit systems.

The $\alpha$-th power of EOF is monogamous for any $N$-qubit system $\rho_{AB_1\cdots B_N-1}$ \cite{Zhu2014},
\begin{eqnarray}\label{jiu1}
E^{\alpha}_f(\rho_{A|B_1\cdots B_N-1})\geq \sum_{i=1}^{N-1}E^{\alpha}_f(\rho_{AB_i})
\end{eqnarray}
for $\alpha\geq\sqrt{2}$, where $E_{f}(\rho_{A|B_1\cdots B_N-1})$ is the bipartite entanglement with respect to the bipartition $A$ and $B_1\cdots B_N-1$ and $E_{f}(\rho_{AB_i})$ is the entanglement of the reduced density operator $\rho_{AB_i}={\rm Tr}_{AB_{1}\cdots B_{i-1}B_{i+1}\cdots B_{N-1}}(\rho_{AB_1\cdots B_N-1})$ of the joint subsystems $A$ and $B_i$ for $i=1, \cdots, N-1$.

From Eqs.~(\ref{s8}) and (\ref{E}), both EOF and $S^{t}$-entropy entanglement have the same monogamy features. Thus for any $N$-qubit system $\rho_{AB_1\cdots B_N-1}$,  one has
\begin{eqnarray}\label{mono1}
E^{\alpha}_t(\rho_{A|B_1\cdots B_N-1})\geq \sum_{i=1}^{N-1} E^{\alpha}_t(\rho_{AB_i})
\end{eqnarray}
for $\alpha\geq\sqrt{2}$.

By using the inequality $(1+t)^{x}\geq1+(2^{x}-1)t^{x}$ for $0\leq t\leq 1$, $x\geq1$~\cite{JQ}, the relation (\ref{mono1}) can be improved as
\begin{eqnarray}\label{xin1}
&&E^\alpha_t(\rho_{A|B_1B_2\cdots B_{N-1}}) \nonumber \\
&&\geqslant  E^\alpha_t(\rho_{AB_1})+\cdots+(2^{\frac{\alpha}{\sqrt{2}}}-1)^{N-3}E^\alpha_t(\rho_{AB_{N-2}}) \nonumber\\
&&~~~+(2^{\frac{\alpha}{\sqrt{2}}}-1)^{N-2}E^\alpha_t(\rho_{AB_{N-1}}),
\end{eqnarray}
with ${E^{\sqrt{2}}_t(\rho_{AB_i}})\geqslant \sum_{j=i+1}^{N-1}E^{\sqrt{2}}_t(\rho_{AB_j})$ for $i=1, 2, \cdots, N-2$, $\alpha\geq\sqrt{2}$. Similarly by using the inequality $(1+t)^{x}\geq1+(2^{x}-t^{x})t^{x}$ for $0\leq t\leq 1$, $x\geq2$~\cite{TYH}, the relation (\ref{xin1}) can be further improved as
\begin{eqnarray}\label{tao1}
&&E^\alpha_t(\rho_{A|B_1B_2\cdots B_{N-1}}) \nonumber \\
&&\geqslant  E^\alpha_t(\rho_{AB_1})+\sum\limits_{i=2}^{N-1}(\prod_{j=1}^{i-1}M_{j})E^\alpha_t(\rho_{AB_i}),
\end{eqnarray}
with ${E^{\sqrt{2}}_t(\rho_{AB_i}})\geqslant \sum_{k=i+1}^{N-1}E^{\sqrt{2}}_t(\rho_{AB_k})$ for $i=1, 2, \cdots, N-2$,
$M_{j}=2^{\frac{\alpha}{\sqrt{2}}}-\left(\frac{\sum_{k=j+1}^{N-1}E^{\sqrt{2}}_t(\rho_{AB_k})}{E^{\sqrt{2}}_t(\rho_{AB_j})}\right)^{\frac{\alpha}{\sqrt{2}}}$ , for $j=1, 2, \cdots, N-2$, $\alpha\geq2\sqrt{2}$.

In the following, we show that the monogamy inequalities (\ref{mono1}), (\ref{xin1}) and (\ref{tao1}) satisfied by the $S^{t}$-entropy entanglement can be further refined and become even tighter. For convenience, we denote $E_{t\,AB_i}=E_{t}(\rho_{AB_i})$ the
$S^{t}$-entropy entanglement of $\rho_{AB_i}$ and $E_{t\,A|B_1,B_2,\cdots,B_{N-1}}=E_{t}(\rho_{A|B_1 \cdots B_{N-1}})$. We first introduce the following lemmas.

\noindent{[\bf Lemma 1]}. Let $t$ and $x$ be real numbers satisfying $0\leqslant t\leqslant 1$ and $x\geqslant 2$. We have
\begin{equation}\label{bud1}
\begin{aligned}
(1+t)^{x-1}\geqslant1+(x-1)t.
\end{aligned}
\end{equation}
\begin{proof}
Set $h(t,x)=(1+t)^{x-1}-(x-1)t-1$ with $0\leqslant t\leqslant 1$ and $x\geqslant 2$. Since $\frac{\partial h(t,x)}{\partial t}=(x-1)(1+t)^{x-2}-(x-1)=(x-1)[(1+t)^{x-2}-1]\geqslant0$, the function $h(t,x)$ is increasing with respect to $t$. As $0\leqslant t \leqslant 1$, $h(t,x)\geq h(0,x)=0$,  we obtain the inequality (\ref{bud1}).
\end{proof}
\noindent{[\bf Lemma 2]}. Let  $x$ be a real number satisfying $x\geqslant 2$. For any $t$ satisfying $0\leqslant t \leqslant 1$, we have
\begin{equation}\label{bud2}
\begin{aligned}
(1+t)^{x}\geqslant1+t+(2^{x}-2)t^{x}.
\end{aligned}
\end{equation}
\begin{proof}
First we note that the above inequality is trivial for $t=0$. So we prove the case for $t\neq0$. Consider the function $f(t,x)=\frac{(1+t)^{x}-t-1}{t^{x}}$ with $0<t \leqslant 1$,  and $x\geqslant 2$.  By using Lemma 1 we have
\begin{equation}
\begin{aligned}
&\frac{\partial f(t,x)}{\partial t}\\
&=\frac{[x(1+t)^{x-1}-1]t^{x}-x t^{x-1}[(1+t)^{x}-t-1]}{t^{2x}}\\
&=\frac{t^{x-1}[-x(1+t)^{x-1}+(x-1)t+x]}{t^{2x}}\leqslant 0,
\nonumber
\end{aligned}
\end{equation}
since $-x(1+t)^{x-1}+(x-1)t+x \leqslant 0$ for $x\geqslant 2$. Therefore, $f(t,x)$ is a decreasing function of $t$. As $0<t \leqslant 1$, we obtain $f(t,x)\geqslant f(1,x)=2^{x}-2$ and the inequality (\ref{bud2}).
\end{proof}
\noindent{[\bf Lemma 3]}. For any $2\otimes2\otimes2$ mixed state $\rho\in {\cal H}_A\otimes {\cal H}_B\otimes {\cal H}_C$, if $E^{\sqrt{2}}_{t\,AB}\geqslant E^{\sqrt{2}}_{t\,AC}$, we have
\begin{equation}\label{mono2}
E^\alpha_{t\,A|BC}\geqslant \Big(1+\frac{E^{\sqrt{2}}_{t\,AC}}
{E^{\sqrt{2}}_{t\,AB}}\Big) E^\alpha_{t\,AB}+(2^{\frac{\alpha}{\sqrt{2}}}-2)E^\alpha_{t\,AC}
\end{equation}
for all $\alpha\geqslant2\sqrt{2}$.
\begin{proof}
By straightforward calculation, if $E^{\sqrt{2}}_{t\,AB}\geqslant E^{\sqrt{2}}_{t\,AC}$ we have
\begin{eqnarray*}
E^\alpha_{t\,A|BC}&&\geqslant (E^{\sqrt{2}}_{t\,AB}+E^{\sqrt{2}}_{t\,AC})^{\frac{\alpha}{\sqrt{2}}}\\
&&=E^\alpha_{t\,AB}\left(1+\frac{E^{\sqrt{2}}_{t\,AC}}
{E^{\sqrt{2}}_{t\,AB}}\right)^{\frac{\alpha}{\sqrt{2}}} \\
&& \geqslant E^\alpha_{t\,AB}\left[1+\frac{E^{\sqrt{2}}_{t\,AC}}
{E^{\sqrt{2}}_{t\,AB}}+(2^{\frac{\alpha}{\sqrt{2}}}-2)\left(\frac{E^{\sqrt{2}}_{t\,AC}}
{E^{\sqrt{2}}_{t\,AB}}\right)^{\frac{\alpha}{\sqrt{2}}}\right]\\
&&=\Big(1+\frac{E^{\sqrt{2}}_{t\,AC}}
{E^{\sqrt{2}}_{t\,AB}}\Big) E^\alpha_{t\,AB}+(2^{\frac{\alpha}{\sqrt{2}}}-2)E^\alpha_{t\,AC},
\end{eqnarray*}
where the second inequality is due to Eq.~(\ref{bud2}) in Lemma 2.
The lower bound becomes trivially zero when $E_{t\,AB}=0$.
\end{proof}

From Lemma 3, we have the following theorem for multi-qubit quantum systems.

\noindent{[\bf Theorem 1]}.
For any $N$-qubit mixed states, if ${E^{\sqrt{2}}_{t\,AB_i}}\geqslant \sum_{j=i+1}^{N-1}E^{\sqrt{2}}_{t\,AB_j}$ for $i=1, 2, \cdots, N-2$, we have
\begin{eqnarray}\label{mono3}
&&E^\alpha_{t\,A|B_1B_2\cdots B_{N-1}} \nonumber \\
&&~\geqslant  \sum\limits_{i=1}^{N-2}(1+\Omega_{i})\Gamma^{i-1} E^\alpha_{t\,AB_i}+\Gamma^{N-2}E^\alpha_{t\,AB_{N-1}}
\end{eqnarray}
for all $\alpha\geqslant2\sqrt{2}$, where $\Gamma=2^{\frac{\alpha}{\sqrt{2}}}-2$,  $\Omega_{i}=\frac{\sum_{j=i+1}^{N-1}E^{\sqrt{2}}_{t\,AB_j}}{E^{\sqrt{2}}_{t\,AB_i}}$, $i=1, 2, \cdots, N-2$.
\begin{proof}
From the inequality (\ref{mono2}), we have
\begin{eqnarray*}\label{o1}
&&E^{\alpha}_{t\,A|B_1B_2\cdots B_{N-1}}\nonumber\\
&&\geqslant  (1+\Omega_{1})E^{\alpha}_{t\,AB_1}+\Gamma
(\sum\limits_{j=2}^{N-1}E^{\sqrt{2}}_{t\,AB_j})^{\frac{\alpha}{\sqrt{2}}}\nonumber\\
&&\geqslant (1+\Omega_{1})E^{\alpha}_{t\,AB_1}+(1+\Omega_{2})\Gamma E^{\alpha}_{t\,AB_2}
 +\Gamma^{2}(\sum\limits_{j=3}^{N-1}E^{\sqrt{2}}_{t\,AB_j})^{\frac{\alpha}{\sqrt{2}}}\nonumber\\
&& \geqslant \cdots\nonumber\\
&&\geqslant (1+\Omega_{1})E^\alpha_{t\,AB_1}+\cdots+(1+\Omega_{N-2})\Gamma^{N-3}E^\alpha_{t\,AB_{N-2}} \nonumber\\
&&~~~+\Gamma^{N-2}E^\alpha_{t\,AB_{N-1}}
\end{eqnarray*}
for all $\alpha\geqslant2\sqrt{2}$.
\end{proof}

\noindent{[\bf Remark 1]}. Theorem 1 gives a new class of monogamy relations for multi-qubit states, which includes the inequality (\ref{xin1}) as a special case of $N=3$, $E_{t\,AB_1}=E_{t\,AB_2}$ and $\alpha\geq2\sqrt{2}$. From the analysis of the aforementioned findings, we observe that different monogamy relationships are characterized by different inequalities, and the compactness of monogamy relations is exactly the compactness of these inequality relations. Since
$$
\begin{aligned}
(1+t)^{x}&\geq1+t+(2^{x}-2)t^{x}\\&=1+(2^{x}-1)t^{x}+t-t^{x}\\&\geq1+(2^{x}-1)t^{x}
\end{aligned}
$$
for $0\leq t\leq1$ and $x\geq2$, where the last inequality is due to that $t-t^{x}\geq0$, obviously our formula (\ref{mono3}) in Theorem 1 gives a tighter monogamy relation (with larger lower bounds) than the inequalities (\ref{mono1}) and (\ref{xin1}) for $\alpha\geq2\sqrt{2}$.

In order to show our formula (\ref{mono3}) in Theorem 1 is indeed tighter than relation (\ref{tao1}), We need introduce the following lemma.

\begin{figure}[h]
	\centering
	\scalebox{2.0}{\includegraphics[width=3.9cm]{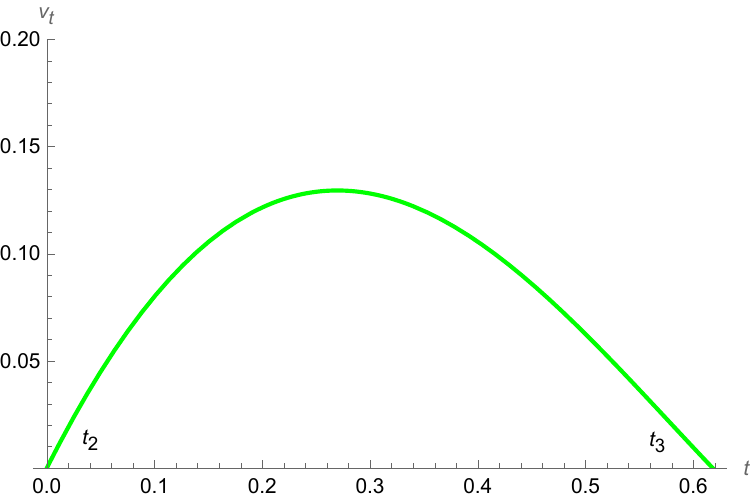}}
	\caption{\small The green line represents the function $v(t)$ with $0\leqslant t\leqslant 1$.}
	\label{Fig1}
\end{figure}

\noindent{[\bf Lemma 4]}. Let $t$ and $x$ be real numbers satisfying $0\leqslant t\leqslant \frac{\sqrt{5}-1}{2}$ and $x\geqslant 2$. We have
\begin{equation}\label{gt}
\begin{aligned}
t-t^{x}\geqslant t^{x}-t^{2x}.
\end{aligned}
\end{equation}
\begin{proof}
Set $u(t,x)=t-2t^{x}+t^{2x}$ with $0\leqslant t\leqslant 1$ and $x\geqslant 2$. Then $\frac{\partial u(t,x)}{\partial x}=-2t^{x}\ln t+2t^{2x}\ln t=2t^{x}\ln t(t^{x}-1)\geqslant0$ as $\ln t\leqslant0$ and $t^{x}-1\leqslant0$. Hence, the function $u(t,x)$ is increasing with respect to $x$. As $ x \geqslant 2$, we get $u(t,x)\geqslant u(t,2)=t-2t^{2}+t^{4}$. Set
$v(t)=t-2t^{2}+t^{4}$. We obtain the four solutions of the equation $v(t)=0$, $t_{1}=\frac{-1-\sqrt{5}}{2}$,  $t_{2}=0$,
$t_{3}=\frac{-1+\sqrt{5}}{2}$ and $t_{4}=1$. Since $v(t)\geqslant 0$ for $0\leqslant t\leqslant \frac{\sqrt{5}-1}{2}$, see Fig.\ref{Fig1}, we have $u(t,x)\geqslant 0$ and obtain the inequality (\ref{gt}).
\end{proof}

\noindent{[\bf Remark 2]}.
In fact, the monogamy relation (\ref{tao1}) is derived from the following inequality,
$$(1+t)^{x}\geq1+(2^{x}-t^{x})t^{x},~ 0\leq t\leq 1,~ x\geq2.$$
Our monogamy relation (\ref{mono3}) is derived from the following inequality,
$$(1+t)^{x}\geqslant1+t+(2^{x}-2)t^{x}, ~ 0\leq t\leq 1,~ x\geq2.$$
Since
$$\begin{aligned}(1+t)^{x}&\geq1+t+(2^{x}-2)t^{x}\\&=1+(2^{x}-1)t^{x}+t-t^{x}\\&\geq1+(2^{x}-1)t^{x}+t^{x}-t^{2x}\\
&=1+(2^{x}-t^{x})t^{x}\end{aligned}$$ for $0\leqslant t\leqslant \frac{\sqrt{5}-1}{2}$ and $x\geqslant 2$, where the second inequality is due to the inequality (\ref{gt}) in Lemma 4, obviously our formula (\ref{mono3}) in Theorem 1 gives a tighter monogamy inequality than (\ref{tao1}) for $\alpha\geq2\sqrt{2}$.

\begin{figure}[h]
	\centering
	\scalebox{2.0}{\includegraphics[width=3.9cm]{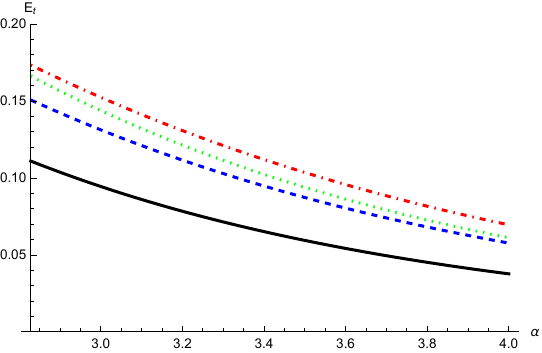}}
	\caption{\small From top to bottom, the red dotdashed  line represents the lower bound from our result (\ref{mono3}) in Theorem 1, the green dotted  line represents the lower bound from the inequality (\ref{tao1}), the blue dashed  line represents the lower bound from the inequality (\ref{xin1}),  the black  line represents the lower bound from the inequality (\ref{mono1}).}
	\label{Fig2}
\end{figure}

\noindent{[\bf Example 1]}. Consider the following three-qubit state $|\psi\rangle$ in generalized Schmidt decomposition \cite{AALE2000,GXH2008},
\begin{eqnarray}
|\psi\rangle_{ABC}
&=&\lambda_0|000\rangle+\lambda_1e^{i\varphi}|100\rangle+\lambda_2|101\rangle
\nonumber\\
&&+\lambda_3|110\rangle
+\lambda_4|111\rangle,
\label{eqn29}
\end{eqnarray}
where $\lambda_i\geq0$, $0\leq\varphi\leq \pi$ and $\sum_{i=0}^4\lambda_i^2=1$. One gets $C(\rho_{A|BC})=2\lambda_0\sqrt{\lambda_2^2+\lambda_3^2+\lambda_4^2}$, $C(\rho_{AB})=2\lambda_0\lambda_2$ and $C(\rho_{AC})=2\lambda_0\lambda_3$. Setting $\lambda_0=\lambda_3=\lambda_4={1}/{\sqrt{5}}$, $\lambda_2=\sqrt{{2}/{5}}$ and $\lambda_1=0$, we have $C(\rho_{A|BC})={4}/{5}$, $C(\rho_{AB})={2\sqrt{2}}/{5}$ and $C(\rho_{AC})={2}/{5}$. By using the equality (\ref{s9}), we obtain the $S^{t}$-entropy entanglement $E_{t\,A|BC}=0.7219$, $E_{t\,AB}=0.4287$ and $E_{t\,AC}=0.2502$. It is seen that our formula (\ref{mono3}) in Theorem 1 is tighter than the inequalities (\ref{mono1}), (\ref{xin1}) and (\ref{tao1}), see Fig.\ref{Fig2}.

Generally, we have the following monogamy inequality.

\noindent{[\bf Theorem 2]}.
For any $N$-qubit mixed states, if ${E^{\sqrt{2}}_{t\,AB_i}}\geqslant \sum_{k=i+1}^{N-1}E^{\sqrt{2}}_{t\,AB_k}$ for $i=1, 2, \cdots, m$, and
${E^{\sqrt{2}}_{t\,AB_j}}\leqslant \sum_{k=j+1}^{N-1}E^{\sqrt{2}}_{t\,AB_k}$ for $j=m+1,\cdots,N-2$,
$\forall$ $1\leq m\leq N-3$, $N\geqslant 4$, we have
\begin{eqnarray}\label{mono4}
&&E^\alpha_{t\,A|B_1B_2\cdots B_{N-1}} \nonumber \\
&&\geqslant  \sum_{i=1}^{m} \Gamma^{i-1}(1+\Omega_{i})E^\alpha_{t\,AB_{i}}+\Gamma^{m+1}
E^\alpha_{t\,AB_{m+1}} \nonumber\\
&&~~~+\Gamma^{m+1}\sum_{j=m+2}^{N-2}(1+\Upsilon_{m+1})\cdots (1+\Upsilon_{j-1})E^\alpha_{t\,AB_{j}}\nonumber\\
&&~~~+\Gamma^{m}(1+\Upsilon_{m+1})\cdots (1+\Upsilon_{N-2})E^\alpha_{t\,AB_{N-1}}
\end{eqnarray}
for all $\alpha\geqslant2\sqrt{2}$, where $\Gamma=2^{\frac{\alpha}{\sqrt{2}}}-2$,  $\Omega_{i}=\frac{\sum_{k=i+1}^{N-1}E^{\sqrt{2}}_{t\,AB_k}}{E^{\sqrt{2}}_{t\,AB_i}}$, $i=1, 2, \cdots, m$,
$\Upsilon_{j}=\frac{E^{\sqrt{2}}_{t\,AB_j}}{\sum_{k=j+1}^{N-1}E^{\sqrt{2}}_{t\,AB_k}}$, $j=m+1, m+2, \cdots, N-2$.

\begin{proof}
From the inequality (\ref{mono2}) in Lemma 3, we have
\begin{eqnarray}\label{j1}
&&E^{\alpha}_{t\,A|B_1B_2\cdots B_{N-1}}\nonumber\\
&&\geqslant  (1+\Omega_{1})E^{\alpha}_{t\,AB_1}+\Gamma
(\sum\limits_{k=2}^{N-1}E^{\sqrt{2}}_{t\,AB_k})^{\frac{\alpha}{\sqrt{2}}}\nonumber\\
&&\geqslant (1+\Omega_{1})E^{\alpha}_{t\,AB_1}+(1+\Omega_{2})\Gamma E^{\alpha}_{t\,AB_2}
 +\Gamma^{2}(\sum\limits_{k=3}^{N-1}E^{\sqrt{2}}_{t\,AB_k})^{\frac{\alpha}{\sqrt{2}}}\nonumber\\
&& \geqslant \cdots\nonumber\\
&&\geqslant (1+\Omega_{1})E^\alpha_{t\,AB_1}+\cdots+(1+\Omega_{m})\Gamma^{m-1}E^\alpha_{t\,AB_{m}} \nonumber\\
&&~~~+\Gamma^{m}(\sum\limits_{k=m+1}^{N-1}E^{\sqrt{2}}_{t\,AB_k})^{\frac{\alpha}{\sqrt{2}}}.
\end{eqnarray}
Similarly, as ${E^{\sqrt{2}}_{t\,AB_j}}\leq \sum\limits_{k=j+1}^{N-1}E^{\sqrt{2}}_{t\,AB_k}$ for $j=m+1,\cdots,N-2$, we get
\begin{eqnarray}\label{j2}
&& (\sum_{k=m+1}^{N-1}E^{\sqrt{2}}_{t\,AB_k})^{\frac{\alpha}{\sqrt{2}}}\nonumber\\
&&\geqslant \Gamma E^{\alpha}_{t\,AB_{m+1}}+(1+\Upsilon_{m+1})(\sum_{k=m+2}^{N-1}E^{\sqrt{2}}_{t\,AB_k})^{\frac{\alpha}{\sqrt{2}}}\nonumber\\
&& \geqslant \cdots\nonumber\\
&&\geqslant \Gamma(E^{\alpha}_{t\,AB_{m+1}}+\cdots+ (1+\Upsilon_{m+1})\cdots (1+\Upsilon_{N-3})E^{\alpha}_{t\,AB_{N-2}})\nonumber\\
&&~~~~+(1+\Upsilon_{m+1})\cdots (1+\Upsilon_{N-2})E^{\alpha}_{t\,AB_{N-1}}.
\end{eqnarray}
Combining Eqs.~(\ref{j1}) and (\ref{j2}), we have Theorem 2.
\end{proof}

Theorem 2 gives another monogamy relation based on the $S^{t}$-entropy entanglement. Comparing inequality (\ref{mono3}) in Theorem 1 with inequality (\ref{mono4}) in Theorem 2,  it is important to point out that for some states that do not meet the conditions outlined in Theorem 1, Theorem 2 may be more effective.

\noindent{[\bf Example 2]}. Consider an $N$-qubit Dicke state \cite{Karmakar(2016)} with $k$ excitations,
\begin{eqnarray}
|D^{(k)}_n\rangle_{A_1A_2\cdots A_n}=\frac{1}{\sqrt{\binom{n}{k}}}\sum_{perm}(|0\rangle^{\otimes (n-k)}|1\rangle^{\otimes k}),
\end{eqnarray}
where the summation is over all possible permutations of the product states having $N-k$ zeros and $k$ ones, and $\binom{N}{k}$ denote the combination number choosing $k$ items from $N$ items.

\begin{figure}[h]
	\centering
	\scalebox{2.0}{\includegraphics[width=3.9cm]{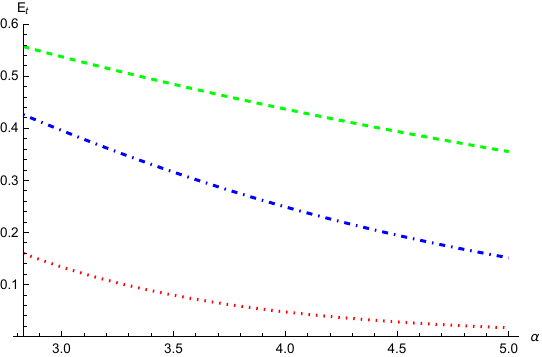}}
	\caption{\small From top to bottom, the green dashed line is the exact values of  $E_{t}(|D^{(1)}_4\rangle_{A_1|A_2A_3A_4})$, the  blue dotdashed  line represents the lower bound from our results (\ref{mono4}) in Theorem 2, the red dotted  line represents the lower bound from the inequality (\ref{mono1}).}
	\label{Fig3}
\end{figure}

The concurrences for Dicke state are given by
\begin{eqnarray}
C(|D^{(k)}_n\rangle_{A_1|A_2\cdots A_n})&=&\frac{2\sqrt{k(n-k)}}{n},
\nonumber\\
C(|D^{(k)}_n\rangle_{A_1A_i})&=&
-\frac{2\sqrt{k(k-1)(n-k)(n-k-1)}}{n(n-1)}
\nonumber
\\
&&+ \frac{2k(n-k)}{n(n-1)},
\end{eqnarray}
where $i\in\{2, \cdots, n\}$. Consider $N=4$ and $k=1$. We get
\begin{eqnarray}
&&C(|D^{(1)}_4\rangle_{A_1|A_2A_3A_4})=\frac{\sqrt{3}}{2},
\nonumber\\
&&C(|D^{(1)}_4\rangle_{A_1A_i})=\frac{1}{2}, i\in\{2, 3, 4\}.
\nonumber\\
\label{eqnDicke3}
\end{eqnarray}
By using equality (\ref{s9}), we get $S^{t}$-entropy entanglement $E_{t}(|D^{(1)}_4\rangle_{A_1|A_2A_3A_4})=0.8113$, $E_{t}(|D^{(1)}_4\rangle_{A_1A_i})=0.3546, i\in\{2, 3, 4\}$.  It is easy to
see that $E_{t}(|D^{(1)}_4\rangle_{A_1A_i})$ does not satisfy the condition (\ref{mono3}) of Theorem 1. From the inequality (\ref{mono4}) of Theorem 2, we have
$$
E_{t}^{\alpha}(|D^{(1)}_4\rangle_{A_1|A_2A_3A_4})\geq(\frac{5}{2}\times2^{\frac{\alpha}{\sqrt{2}}}-2)(0.3546)^{\alpha}
$$
for $\alpha\geq2\sqrt{2}$, see Fig.\ref{Fig3}.

\section{Monogamy of $T^{t}_q$-entropy entanglement}\label{Sec2}
For a pure state $|\Phi\rangle_{AB}$ on Hilbert space ${\cal H}_A\otimes {\cal H}_{B}$, the $T^{t}_q$-entropy entanglement is defined by
\begin{eqnarray}
\mathcal{T}^{t}_q(|\Phi\rangle_{AB})=T^t_q(\rho_A),
\label{t1}
\end{eqnarray}
where $\rho_A={\rm Tr}_B(|\Phi\rangle_{AB}\langle\Phi|)$ denotes the reduced density operator of the subsystem $A$.
$T^t_q(\rho)$ is the total entropy of the Tsallis-$q$ entropy. Its complementary dual of a quantum state $\rho$ on $d$-dimensional Hilbert space ${\cal H}$ is defined by
\begin{eqnarray}
T^{t}_q(\rho)&=&\frac{1-{\rm Tr}\rho^q-{\rm Tr}({\mathbb{1}-\rho})^q+{\rm Tr}({\mathbb{1}-\rho)}}{q-1}.
\label{t2}
\end{eqnarray}
For a bipartite mixed state $\rho_{AB}$ on Hilbert space ${\cal H}_A\otimes {\cal H}_{B}$, the $T^{t}_q$-entropy entanglement is defined via convex-roof extension
\begin{eqnarray}
\mathcal{T}^{t}_q(\rho_{AB})=\inf_{\{p_i,|\Phi_i\rangle\}}\sum_ip_i\mathcal{T}^{t}_q(|\Phi_i\rangle_{AB}),
\label{t3}
\end{eqnarray}
where the infimum is taken over all the possible pure-state decompositions of $\rho_{AB}=\sum_ip_i|\Phi_i\rangle_{AB}\langle\Phi_i|$.

For a bipartite pure state $|\psi\rangle_{AB}$, the Tsallis-$q$ entanglement is defined by \cite{Kim2010T}
\begin{eqnarray}\label{t4}
T_q(|\psi\rangle_{AB})=S_q(\rho_A)=\frac{1}{q-1}(1-\mathrm{tr}\rho_A^q),
\end{eqnarray}
for any $q > 0$ and $q \ne 1$. If $q$ tends to 1, $T_q(\rho)$ converges to the von Neumann entropy, $\lim_{q\to1} T_q(\rho)=-\mathrm{tr}\rho\ln\rho=S_q(\rho)$. For a bipartite mixed state $\rho_{AB}$, Tsallis-$q$ entanglement is defined via the convex-roof extension,
$$
T_q(\rho_{AB})=\min\sum_ip_iT_q(|\psi_i\rangle_{AB}),
$$
with the minimum taken over all possible pure-state decompositions of $\rho_{AB}$.

In Ref.~\cite{YGM2016} the authors presented an analytic relation between Tsallis-$q$ entanglement and concurrence for $\frac{5-\sqrt{13}}{2}\leq q\leq \frac{5+\sqrt{13}}{2}$,
\begin{eqnarray}\label{an1}
T_q(|\psi\rangle_{AB})=\textsl{g}_q(C^2(|\psi\rangle_{AB})),
\end{eqnarray}
where the function $\textsl{g}_q(x)$ is defined as
\small
\begin{eqnarray}\label{an2}
\textsl{g}_q(x)=\frac{\left[1-\left(\frac{1+\sqrt{1-x}}{2}\right)^q-\left(\frac{1-\sqrt{1-x}}{2}\right)^q\right]}{q-1}.
\end{eqnarray}
\normalsize
It has been shown that $T_q(|\psi\rangle)=\textsl{g}_q\left(C^2(|\psi\rangle)\right)$ for any $2\otimes m~(m\geqslant2)$ pure state $|\psi\rangle$, and  $T_q(\rho)=\textsl{g}_q\left(C^2(\rho)\right)$ for any two-qubit mixed state $\rho$ in Ref.~\cite{Kim2010T}. For any $N$-qubit system $\rho_{AB_2\cdots B_N-1}$, it is further proved that
\begin{eqnarray}\label{t55}
{T_q^\beta}(\rho_{A|B_1B_2\cdots B_{N-1}})\geqslant\sum_{i=1}^{N-1}{T_q^\beta}(\rho_{AB_i}),
\end{eqnarray}
with $\frac{5-\sqrt{13}}{2}\leqslant q\leqslant \frac{5+\sqrt{13}}{2}$ and $\beta\geq2$.

Consider an arbitrary pure state $\phi_{AB}$ given by Eq.~(\ref{Schmidt}). It can be verified that
\begin{eqnarray}
\mathcal{T}^{t}_q(|\phi\rangle_{AB})=f_q(C(|\phi\rangle_{AB})),
\label{t4}
\end{eqnarray}
where the analytic function $f_q(x)$ is defined by
\begin{eqnarray}\label{an22}
f_q(x)&=&\frac{2[1-(\frac{1+\sqrt{1-x^2}}{2})^q-(\frac{1-\sqrt{1-x^2}}{2})^q]}{q-1}.
\end{eqnarray}
 From Ref.~\cite{Kim2010T} we have the functional relation,
\begin{eqnarray}
\mathcal{T}^{t}_q(\rho_{AB})=f_q(C(\rho_{AB}))
\label{t5}
\end{eqnarray}
for a bipartite two-qubit mixed state $\rho_{AB}$ on Hilbert space $\mathcal{H}_A\otimes\mathcal{H}_B$. From Eqs.~(\ref{an2}) and (\ref{an22}), this means that the $T^{t}_q$-entropy entanglement for qubit systems can be reduced to the Tsallis-$q$ entropy entanglement. Thus, a general monogamy of the Tsallis-$q$ entropy entanglement in multi-qubit systems is naturally inherited by the $T^{t}_q$-entropy entanglement \cite{Kim2010T,Luo2016}.
For any $N$-qubit system $\rho_{AB_2\cdots B_N-1}$, we have
\begin{eqnarray}\label{mono5}
(\mathcal{T}^{t}_{q})^{\beta}(\rho_{A|B_1\cdots B_N-1})\geq \sum_{i=1}^{N-1} (\mathcal{T}^{t}_{q})^{\beta}(\rho_{A|B_i}),
\end{eqnarray}
with $\frac{5-\sqrt{13}}{2}\leqslant q\leqslant \frac{5+\sqrt{13}}{2}$ and $\beta\geq2$.

By using the inequality $(1+t)^{x}\geq1+(2^{x}-1)t^{x}$ for $0\leq t\leq 1$ and $x\geq1$~\cite{JQ},
the relation (\ref{mono5}) is improved for $\beta\geq2$ as
\begin{eqnarray}\label{xin2}
&&(\mathcal{T}^{t}_{q})^{\beta}(\rho_{A|B_1B_2\cdots B_{N-1}}) \nonumber \\
&&\geqslant  (\mathcal{T}^{t}_{q})^{\beta}(\rho_{AB_1})+\cdots+(2^{\frac{\beta}{2}}-1)^{N-3}(\mathcal{T}^{t}_{q})^{\beta}(\rho_{AB_{N-2}}) \nonumber\\
&&~~~+(2^{\frac{\beta}{2}}-1)^{N-2}(\mathcal{T}^{t}_{q})^{\beta}(\rho_{AB_{N-1}})
\end{eqnarray}
with $({\mathcal{T}^{t}_{q})^{2}(\rho_{AB_i}})\geqslant \sum_{j=i+1}^{N-1}(\mathcal{T}^{t}_{q})^{2}(\rho_{AB_j})$ for $i=1, 2, \cdots, N-2$, $\frac{5-\sqrt{13}}{2}\leqslant q\leqslant \frac{5+\sqrt{13}}{2}$. Similarly by using the inequality $(1+t)^{x}\geq1+(2^{x}-t^{x})t^{x}$ for $0\leq t\leq 1$ and $x\geq2$~\cite{TYH}, the relation (\ref{xin2}) is further improved as
\begin{eqnarray}\label{tao2}
&&(\mathcal{T}^{t}_{q})^{\beta}(\rho_{A|B_1B_2\cdots B_{N-1}}) \nonumber \\
&&\geqslant  (\mathcal{T}^{t}_{q})^{\beta}(\rho_{AB_1})+\sum\limits_{i=2}^{N-1}(\prod_{j=1}^{i-1}M_{j})(\mathcal{T}^{t}_{q})^{\beta}(\rho_{AB_i})
\end{eqnarray}
with ${(\mathcal{T}^{t}_{q})^{2}(\rho_{AB_i}})\geqslant \sum_{k=i+1}^{N-1}(\mathcal{T}^{t}_{q})^{\beta}(\rho_{AB_k})$ for $i=1, 2, \cdots, N-2$,
$M_{j}=2^{\frac{\beta}{2}}-\left(\frac{\sum_{k=j+1}^{N-1}(\mathcal{T}^{t}_{q})^{2}(\rho_{AB_k})}{(\mathcal{T}^{t}_{q})^{2}(\rho_{AB_j})}\right)^{\frac{\beta}{2}}$ , for $j=1, 2, \cdots, N-2$, $\frac{5-\sqrt{13}}{2}\leqslant q\leqslant \frac{5+\sqrt{13}}{2}$ and $\beta\geq4$.

In the following, we show that these monogamy inequalities satisfied by the $T^{t}_q$-entropy entanglement can be further refined and become even tighter. For convenience, we denote $T_{AB_i}=\mathcal{T}^{t}_{q}(\rho_{AB_i})$ the $T^{t}_q$-entropy entanglement of $\rho_{AB_i}$ and $T_{A|B_1,B_2,\cdots,B_{N-1}}=\mathcal{T}^{t}_{q}(\rho_{A|B_1 \cdots B_{N-1}})$. We first introduce a lemma.

\noindent{[\bf Lemma 5]}. For any $2\otimes2\otimes2$ mixed state $\rho\in {\cal H}_A\otimes {\cal H}_B\otimes {\cal H}_C$, if $T^{2}_{AB}\geqslant T^{2}_{AC}$, we have
\begin{equation}\label{mono6}
  T^\beta_{A|BC}\geqslant \left(1+\frac{T^{2}_{AC}}{T^{2}_{AB}}\right) T^\beta_{AB}+(2^{\frac{\beta}{2}}-2)T^\beta_{AC}
\end{equation}
for all $\beta\geqslant4$ and $\frac{5-\sqrt{13}}{2}\leqslant q\leqslant \frac{5+\sqrt{13}}{2}$.

\begin{proof}
By straightforward calculation, if $T^{2}_{AB}\geqslant T^{2}_{AC}$ we have
\begin{eqnarray*}
  T^{\beta}_{A|BC}&&\geqslant (T^{2}_{AB}+T^{2}_{AC})^{\frac{\beta}{2}}\\
  &&=T^{\beta}_{AB}\left(1+\frac{T^{2}_{AC}}{T^{2}_{AB}}\right)^{\frac{\beta}{2}} \\
  && \geqslant T^{\beta}_{AB}\left[1+\frac{T^{2}_{AC}}{T^{2}_{AB}}+(2^{\frac{\beta}{2}}-2)\left(\frac{T^{2}_{AC}}{T^{2}_{AB}}\right)^{\frac{\beta}{2}}\right]\\
  &&= \left(1+\frac{T^{2}_{AC}}{T^{2}_{AB}}\right) T^\beta_{AB}+(2^{\frac{\beta}{2}}-2)T^\beta_{AC},
\end{eqnarray*}
where the second inequality is due to Lemma 2. As the subsystems $A$ and $B$ are equivalent in this case, we have assumed that $T_{AB}\geqslant T_{AC}$ without loss of generality. Moreover,
if $T_{AB}=0$ we have $T_{AB}=T_{AC}=0$. That is to say the lower bound becomes trivially zero.
\end{proof}

From Lemma 5, we have the following theorem.

\noindent{[\bf Theorem 3]}.
For any $N$-qubit mixed state, if ${T^{2}_{AB_i}}\geqslant \sum_{j=i+1}^{N-1}T^{2}_{AB_j}$ for $i=1, 2, \cdots, N-2$, we have
\begin{eqnarray}\label{mono7}
&&T^{\beta}_{A|B_1B_2\cdots B_{N-1}} \nonumber \\
&&~\geqslant  \sum\limits_{i=1}^{N-2}(1+\Omega_{i})\Gamma^{i-1} T^{\beta}_{AB_i}+\Gamma^{N-2}T^{\beta}_{AB_{N-1}}
\end{eqnarray}
for all $\beta\geqslant4$ and $\frac{5-\sqrt{13}}{2}\leqslant q\leqslant \frac{5+\sqrt{13}}{2}$, where $\Gamma=2^{\frac{\beta}{2}}-2$,  $\Omega_{i}=\frac{\sum_{j=i+1}^{N-1}T^{2}_{AB_j}}{T^{2}_{AB_i}}$, $i=1, 2, \cdots, N-2$.

\begin{proof}
From the inequality (\ref{mono6}) in Lemma 5, we have
\begin{eqnarray}
&&T^{\beta}_{A|B_1B_2\cdots B_{N-1}}\nonumber\\
&&\geqslant  (1+\Omega_{1})T^{\beta}_{AB_1}+\Gamma
(\sum\limits_{j=2}^{N-1}T^{2}_{AB_j})^{\frac{\beta}{2}}\nonumber\\
&&\geqslant (1+\Omega_{1})T^{\beta}_{AB_1}+(1+\Omega_{2})\Gamma T^{\beta}_{AB_2}
 +\Gamma^{2}(\sum\limits_{j=3}^{N-1}T^{2}_{AB_j})^{\frac{\beta}{2}}\nonumber\\
&& \geqslant \cdots\nonumber\\
&&\geqslant (1+\Omega_{1})T^{\beta}_{AB_1}+\cdots+(1+\Omega_{N-2})\Gamma^{N-3}T^{\beta}_{AB_{N-2}} \nonumber\\
&&~~~+\Gamma^{N-2}T^{\beta}_{AB_{N-1}}
\end{eqnarray}
for all $\beta\geqslant4$ and $\frac{5-\sqrt{13}}{2}\leqslant q\leqslant \frac{5+\sqrt{13}}{2}$.
\end{proof}

\noindent{[\bf Remark 3]}. Theorem 3 introduces a new class of monogamy relations for multi-qubit states, encompassing inequality (\ref{xin2}) as a specific case of $N=3$, $T_{AB_{1}}=T_{AB_{2}}$ and $\beta\geq4$. Similar to the discussions in Remark 1 and  Remark 2, our formula (\ref{mono7}) in Theorem 3 gives a tighter monogamy relation with larger lower bounds than the inequalities (\ref{mono5}), (\ref{xin2}) and (\ref{tao2}).

\begin{figure}[h]
\centering
\scalebox{2.0}{\includegraphics[width=3.9cm]{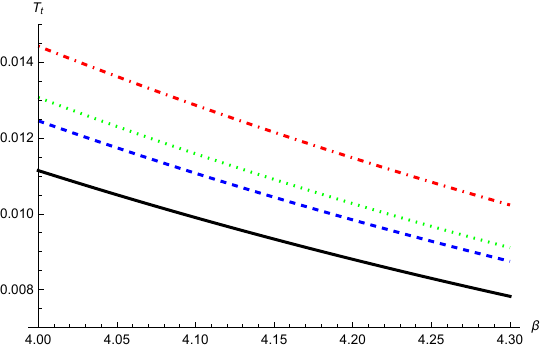}}
\caption{\small From top to bottom, the  red dotdashed line represents the lower bound from our result (\ref{mono7}) in Theorem 3, the green dotted line represents the lower bound from the inequality (\ref{tao2}), the blue dashed  line represents the lower bound from the inequality (\ref{xin2}), the black  line represents the lower bound from the inequality (\ref{mono5}).
}
\label{Fig4}
\end{figure}

\noindent{[\bf Example 3]}. Let us again consider the three-qubit state $|\psi\rangle_{ABC}$ in Example 1. Setting $\lambda_0=\lambda_3=\lambda_4={1}/{\sqrt{5}}$, $\lambda_2=\sqrt{{2}/{5}}$ and $\lambda_1=0$, we have $C(\rho_{A|BC})={4}/{5}$, $C(\rho_{AB})={2\sqrt{2}}/{5}$ and $C(\rho_{AC})={2}/{5}$.
By using equality (\ref{t5}) and taking $q=2$, we get the $T^{t}_{q}$-entropy entanglement of $|\psi\rangle_{ABC}$, $T_{A|BC}=0.64$, $T_{AB}=0.32$ and $T_{AC}=0.16$.
It is seen that our formula (\ref{mono7}) in Theorem 3 is tighter than inequalities (\ref{mono5}), (\ref{xin2}) and (\ref{tao2}) for $\beta\geq4$, see Fig.\ref{Fig4}.

Generally, the conditions for inequalities (\ref{mono7}) are not always satisfied. In following, we present a general
monogamy inequality.

\noindent{[\bf Theorem 4]}.
For an $N$-qubit mixed state, if ${T^{2}_{AB_i}}\geqslant \sum_{k=i+1}^{N-1}T^{2}_{AB_k}$ for $i=1, 2, \cdots, m$, and
${T^{2}_{AB_j}}\leqslant \sum_{k=j+1}^{N-1}T^{2}_{AB_k}$ for $j=m+1,\cdots,N-2$,
$\forall$ $1\leq m\leq N-3$, $N\geqslant 4$, we have
\begin{eqnarray}\label{mono8}
&&T^{\beta}_{A|B_1B_2\cdots B_{N-1}} \nonumber \\
&&\geqslant  \sum_{i=1}^{m} \Gamma^{i-1}(1+\Omega_{i})T^{\beta}_{AB_{i}}+\Gamma^{m+1}
T^{\beta}_{AB_{m+1}} \nonumber\\
&&~~~+\Gamma^{m+1}\sum_{j=m+2}^{N-2}(1+\Upsilon_{m+1})\cdots (1+\Upsilon_{j-1})T^{\beta}_{AB_{j}}\nonumber\\
&&~~~+\Gamma^{m}(1+\Upsilon_{m+1})\cdots (1+\Upsilon_{N-2})T^{\beta}_{AB_{N-1}}
\end{eqnarray}
for all $\beta\geqslant4$ and $\frac{5-\sqrt{13}}{2}\leqslant q\leqslant \frac{5+\sqrt{13}}{2}$,
$\Omega_{i}=\frac{\sum_{k=i+1}^{N-1}T^{2}_{AB_k}}{T^{2}_{AB_i}}$, $i=1, 2, \cdots, m$,
$\Upsilon_{j}=\frac{T^{2}_{AB_j}}{\sum_{k=j+1}^{N-1}T^{2}_{AB_k}}$, $j=m+1, m+2, \cdots, N-2$.

\begin{proof}
From the inequality (\ref{mono6}) in Lemma 5, we have
\begin{eqnarray}\label{j3}
&&T^{\beta}_{A|B_1B_2\cdots B_{N-1}}\nonumber\\
&&\geqslant  (1+\Omega_{1})T^{\beta}_{AB_1}+\Gamma
(\sum\limits_{k=2}^{N-1}T^{2}_{AB_k})^{\frac{\beta}{2}}\nonumber\\
&&\geqslant (1+\Omega_{1})T^{\beta}_{AB_1}+(1+\Omega_{2})\Gamma T^{\beta}_{AB_2}
 +\Gamma^{2}(\sum\limits_{k=3}^{N-1}T^{2}_{AB_k})^{\frac{\beta}{2}}\nonumber\\
&& \geqslant \cdots\nonumber\\
&&\geqslant (1+\Omega_{1})T^{\beta}_{AB_1}+\cdots+(1+\Omega_{m})\Gamma^{m-1}T^{\beta}_{AB_{m}} \nonumber\\
&&~~~+\Gamma^{m}(\sum\limits_{k=m+1}^{N-1}T^{2}_{AB_k})^{\frac{\beta}{2}}.
\end{eqnarray}
Similarly, as ${T^{2}_{AB_j}}\leq \sum\limits_{k=j+1}^{N-1}T^{2}_{AB_k}$ for $j=m+1,\cdots,N-2$, we get
\begin{eqnarray}\label{j4}
&& (\sum_{k=m+1}^{N-1}T^{2}_{AB_k})^{\frac{\beta}{2}}\nonumber\\
&&\geqslant \Gamma T^{\beta}_{AB_{m+1}}+(1+\Upsilon_{m+1})(\sum_{k=m+2}^{N-1}T^{2}_{AB_k})^{\frac{\beta}{2}}\nonumber\\
&& \geqslant \cdots\nonumber\\
&&\geqslant \Gamma(T^{\beta}_{AB_{m+1}}+\cdots+ (1+\Upsilon_{m+1})\cdots (1+\Upsilon_{N-3})T^{\beta}_{AB_{N-2}})\nonumber\\
&&~~~~+(1+\Upsilon_{m+1})\cdots (1+\Upsilon_{N-2})T^{\beta}_{AB_{N-1}}.
\end{eqnarray}
Combining Eqs.~(\ref{j3}) and (\ref{j4}), we have Theorem 4.
\end{proof}

Theorem 4 gives another monogamy relation based on the $T^{t}_{q}$-entropy entanglement.  Comparing inequality (\ref{mono7}) in Theorem 3 with inequality (\ref{mono8}) in Theorem 4, one notices that for some classes of states that do not satisfy the conditions in Theorem 3, Theorem 4 works still.

\begin{figure}[h]
	\centering
	\scalebox{2.0}{\includegraphics[width=3.9cm]{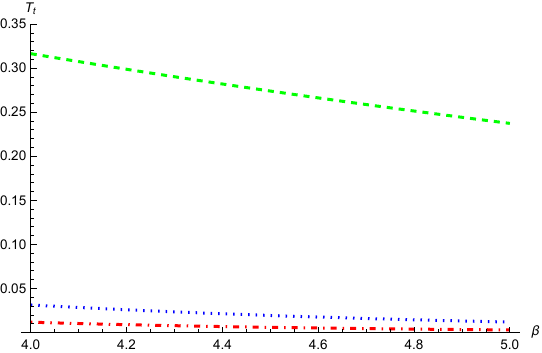}}
	\caption{\small  From top to bottom, the green dashed line is the exact values of  $T_{A|BCD}$, the  blue dotted  line represents the lower bound from our results (\ref{mono8}) in Theorem 4, the red  dotdashed line represents the lower bound from the inequality (\ref{mono5}).
}
	\label{Fig5}
\end{figure}

\noindent{[\bf Example 4]}.  Let us consider the four-qubit generalized $W$ state,
$$
|W\rangle_{ABCD}=\frac{1}{2}(|1000\rangle+|0100\rangle+|0010\rangle+|0001\rangle).
$$
Suppose $q = 2$. We have $T_{A|BCD}={3}/{4}$ and $T_{AB}=T_{AC}=T_{AD}=\frac{1}{4}$.
It is easy to see that this state does not satisfy the condition (\ref{mono7}) of Theorem 3. From inequality (\ref{mono8}) of Theorem 4, we have
$$
T^{\beta}_{A|BCD}\geq(\frac{5}{2}\times2^{\frac{\beta}{2}}-2)(\frac{1}{4})^{\beta}
$$
for $\beta\geq4$, see Fig.\ref{Fig5}.

\section{ Two NEW KINDs OF MULTIPARTITE ENTANGLEMENT INDICATORS}\label{Sec3}
Based on the monogamy relations (\ref{mono1}) and (\ref{mono5}), we are able to construct two sets of useful entanglement indicators that can be utilized to identify all genuine multi-qubit entangled states even for the cases that the three tangle of concurrence does not work. Let us first recall the definition of tangle.
The tangle of a bipartite pure states $|\psi\rangle_{AB}$ is defined as \cite{CKW2000},
\begin{equation}
\tau(|\psi\rangle_{AB})=2(1-{\rm tr}\rho_A^2),
\end{equation}
where $\rho_A={\rm tr}_B|\psi\rangle_{AB}\langle\psi|$.
The tangle of a bipartite mixed state $\rho_{AB}$ is defined as
\begin{equation}\label{tauAB}
\tau(\rho_{AB})=\Bigg[\min\limits_{\{p_k,|\psi_k\rangle\}}\sum\limits_{k}p_k\sqrt{\tau(|\psi_k\rangle_{AB})}\Bigg]^2,
\end{equation}
where the minimization in Eq.~\eqref{tauAB} is taken over all possible pure state decompositions of $\rho_{AB}=\sum\nolimits_{k}p_k|\psi_k\rangle_{AB}\langle\psi_k|$.

Based on Eq.~(\ref{mono1}), we can construct a class of multipartite entanglement indicators in terms of the $S_{t}$-entropy entanglement,
\begin{equation}\label{d1}
\tau_t(\rho_{A|B_1\ldots{B_N-1}})=\min\sum_ip_i\tau_t(|\psi_{A|B_1\ldots{B_N-1}}^i\rangle),
\end{equation}
where the minimum is taken over all possible pure state decompositions $\{p_i,\psi_{A|B_1\ldots{B_N-1}}^i\}$ of $\rho_{AB_1\ldots{B_N-1}}$ and $\tau_t(|\psi_{A|B_1\ldots{B_N-1}}^i\rangle=E_{t}^{\sqrt{2}}(\psi_{A|B_1\ldots{B_N-1}}^i)-\sum_{j=1}^{N-1} E_{t}^{\sqrt{2}}(\rho_{AB_{j}}^i)$.

Similarly, based on Eq.~(\ref{mono5}), we can construct a class of multipartite entanglement indicators in terms of the $T^{t}_{q}$-entropy entanglement for $\frac{5-\sqrt{13}}{2}\leqslant q\leqslant \frac{5+\sqrt{13}}{2}$,
\begin{equation}\label{d2}
\omega_q(\rho_{A|B_1\ldots{B_N-1}})=\min\sum_ip_i\omega_q(|\psi_{A|B_1\ldots{B_N-1}}^i\rangle),
\end{equation}
where the minimum is taken over all possible pure state decompositions $\{p_i,\psi_{A|B_1\ldots{B_N-1}}^i\}$ of $\rho_{AB_1\ldots{B_N-1}}$ and $\omega_q(|\psi_{A|B_1\ldots{B_N-1}}^i\rangle=(\mathcal{T}_{q}^{t})^{2}(\psi_{A|B_1\ldots{B_N-1}}^i)-\sum_{j=1}^{N-1} (\mathcal{T}_{q}^{t})^{2}(\rho_{AB_{j}}^i)$.

In particular, we evaluate Eqs.~(\ref{d1}) and (\ref{d2}) for the $W$-state.  The nonzero values of  $\tau_t$ and $\omega_q$ in following example assert their validity as two genuine entanglement indicators.

\begin{figure}[h]
	\centering
	\scalebox{2}{\includegraphics[width=3.9cm]{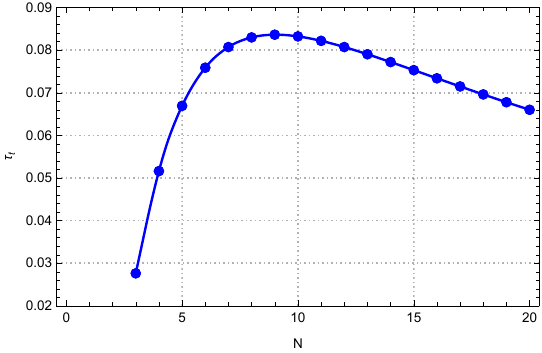}}
	\caption{\small  The nonzero
values indicate the existence of genuine multipartite entanglement in an $N$-qubit $W$ state.}
	\label{Fig6}
\end{figure}

\noindent{[\bf Example 5]}. We consider the $N$-qubit $W$ state,
$$
|W\rangle_N=\frac{1}{\sqrt N}(|10\cdots0\rangle+|01\cdots0\rangle+|0\cdots01\rangle).
$$
The three tangle cannot detect the genuine tripartite entanglement of the $W$-state. However, the indicator $\tau_{t}$ works in this case. By using the multipartite entanglement indicator given in Eq.~(\ref{d1}), we have $\tau_{t}(|W\rangle_N)=h^{\sqrt{2}}(\frac{2\sqrt{N-1}}{N})-(N-1)h^{\sqrt{2}}(\frac{2}{N})$.  We plot the indicator as a function of $N$ in a $N$-qubit $W$ state, where the nonzero values imply that the genuine multipartite entanglement is detected, see Fig.\ref{Fig6}.

\begin{figure}[h]
	\centering
	\scalebox{2}{\includegraphics[width=3.9cm]{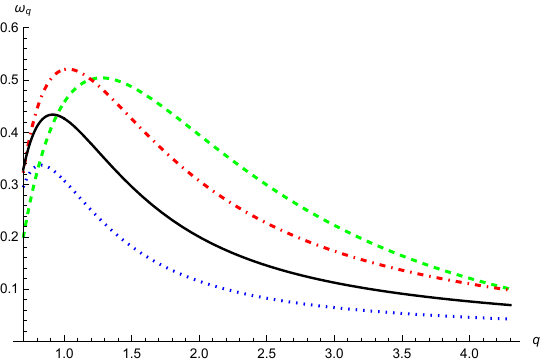}}
	\caption{\small  The  green dashed (red dotdashed, black, blue dotted)  line represents the the value of $\omega_q$ for $N=3 ~(5,~ 7,~10)$, respectively.}
	\label{Fig7}
\end{figure}

Moreover, the indicator $\omega_{q}$ effectively detects the genuine multipartite entanglement in this state too. By using the multipartite entanglement indicator given in Eq.~(\ref{d2}), we have $\omega_q(|W\rangle_N)=f_q^2(\frac{2\sqrt{N-1}}{N})-(N-1)f_q^2(\frac{2}{N})$. We plot the indicator as a function of $q$ for $N=3,5,7,10$, respectively.  It shows that the indicator $\omega_q(|W\rangle)$ is always positive for $q\in[\frac{5-\sqrt{13}}{2},\frac{5+\sqrt{13}}{2}]$, see Fig.\ref{Fig7}.

\section{Conclusion}\label{Sec4}
The monogamy relationship of quantum entanglement embodies fundamental properties manifested by multipartite entangled states. We have provided the general monogamy relations for two new entanglement measures in multi-qubit quantum systems, and demonstrated that these inequalities give rise to tighter constraints than the existing ones. Detailed examples have been presented to illustrate the effectiveness of our results in characterizing the multipartite entanglement distributions. Based on these general monogamy relations, we are able to construct the set of multipartite entanglement indicators for $N$-qubit states, which work well even when the concurrence-based indicators fails to detect the genuine multipartite entanglement. The distribution of entanglement in multipartite  systems can be more precisely characterized through stricter monogamy inequalities. Our results may shed new light on further investigations of comprehending the distribution of entanglement in multipartite systems.

\bigskip
\noindent{\bf Acknowledgments}\, \,
This work is supported by the National Natural Science Foundation of China (NSFC) under Grants 12075159, 12171044 and 12301582; the specific research fund of the Innovation Platform for Academicians of Hainan Province; the Start-up Funding of Dongguan University of Technology No. 221110084.

\end{document}